\documentclass[]{aa}  

\usepackage{graphicx}
\usepackage{ulem}
\usepackage{hyperref}
\usepackage{txfonts}
\usepackage{todonotes}
\usepackage{subcaption}
\usepackage{xcolor}
\usepackage{fancyhdr}

\usepackage[printwatermark]{xwatermark}
\newwatermark[page=1,color=black!100,angle=0,scale=0.32,xpos=77,ypos=142]{NORDITA 2022-092}

\hypersetup{colorlinks=true, linkcolor=blue, citecolor = blue, urlcolor=cyan }

\newcommand{\schXXII}{\hyperlink{cite.schneider_2022_ksz}{Sch22}}
\newcommand{\eXV}{\hyperlink{cite.Enqvist_2015}{E15}}
\newcommand{\eXX}{\hyperlink{cite.enqvist_2020}{E20}}
\newcommand{\sXXII}{\hyperlink{cite.Simon_2022}{S22}}
\newcommand{\aXXI}{\hyperlink{cite.kids_1000}{A21}}
\newcommand{\planck}{\hyperlink{cite.planck_2018}{P20b}}

\begin{document} 

\title{Constraining dark matter decay with cosmic microwave background and weak-lensing shear observations}

  \author{Jozef Bucko\thanks{jozef.bucko@uzh.ch}\inst{1}
          \and
          Sambit K. Giri\inst{1,2}
          \and
          Aurel Schneider\inst{1}
          }

  \institute{Institute for Computational Science, University of Zurich, Winterthurerstrasse 190, 8057 Zurich, Switzerland \and
  Nordita, KTH Royal Institute of Technology and Stockholm University, Hannes Alf\'vens v\"ag 12, SE-106 91 Stockholm, Sweden\\
             }

  \date{Received: 28 November 2022. Accepted: 12  February 2023. 
  }

  \abstract
   {From observations at low and high redshifts, it is well known that the bulk of dark matter (DM) has to be stable or at least very long-lived. 
   However, the possibility that a small fraction of DM is unstable or that all DM decays with a half-life time ($\tau$) significantly longer than the age of the Universe is not ruled out. One-body decaying dark matter (DDM) consists of a minimal extension to the $\Lambda$CDM model. It causes a modification of the cosmic growth history as well as a suppression of the small-scale clustering signal, providing interesting consequences regarding the $S_8$  tension, which is the observed difference in the clustering amplitude between weak-lensing (WL) and cosmic microwave background (CMB) observations. In this paper, we investigate models in which a fraction or all DM decays into radiation, focusing on the long-lived regime, that is, $\tau \gtrsim H_0^{-1}$ ( $H_0^{-1}$ being the Hubble time). 
   We used WL data from the Kilo-Degree Survey ({\tt KiDS}) and CMB data from {\tt Planck}. First, we confirm that this DDM model cannot alleviate the $S_8$ difference. We then show that the most constraining power for DM decay does not come from the nonlinear WL data, but from CMB via the integrated Sachs-Wolfe effect. From the CMB data alone, we obtain constraints of $\tau \geq 288$~Gyr if all DM is assumed to be unstable, and we show that a maximum fraction of $f=0.07$ is allowed to decay assuming the half-life time to be comparable to (or shorter than) one Hubble time. The constraints from the {\tt KiDS}-1000 WL data are significantly weaker, $\tau \geq 60$~Gyr and $f<0.34$. Combining the CMB and WL data does not yield tighter constraints than the CMB alone, except for short half-life times, for which the maximum allowed fraction becomes $f=0.03$. All limits are provided at the 95\% confidence level.}


   \maketitle

%

\section{Introduction}
\label{sec:intro}

There is overwhelming evidence for the existence of dark matter (DM), but we still know very little about its nature and composition. DM most probably consists of one or several new particles, requiring an extension of the standard model \citep{bertone2005particle,feng2010dark}. The bulk of these particles has to be rather cold, interacts weakly at most, and is stable over at least one Hubble time. However, small deviations from these assumptions remain possible. Furthermore, a multi-particle DM sector would allow sub-species to evade the requirements mentioned above. They might be hot, interact strongly, or be very unstable, for instance.

In this paper, we focus on the possibility that a fraction or all of the DM fluid decays into radiation via a simple one-body decay channel. The nature of this radiation component is not specified and is not relevant to our analysis. Decay into photons or other standard model particles would lead to constraints from the absence of an observable radiation signal in the sky, however, which would exceed the constraints provided here. We therefore implicitly assume a DM decay into dark radiation.

Recent weak-lensing (WL) surveys such as {\tt CFHTLenS}\footnote{Canada-France-Hawaii Telescope Lensing  Survey} \citep{heymans_2021_CFHTLenS,fu_2015_cfhtlens}, {\tt KiDS}\footnote{Kilo-Degree Survey} (\citealt{Kuijken_2019_kids,Giblin_2021_kids_1000_catalogue,Hildebrandt_2021_kids1000_catalogue}; \citealt[A21]{kids_1000}), {\tt HSC}\footnote{Hyper Supreme-Cam} \citep{aihara_2017_hsc,hsc_y1_weak_lensing,HSC_DR_Y3,liu_2022_hsc}, and {\tt DES}\footnote{Dark Energy Survey} \citep{DES_2005,des_y3,amon_2022_des} have reported a mild but persistent difference of the clustering amplitude $\sigma_8$ of the cosmic microwave background (CMB) as measured by the {\tt Planck} satellite \citep{planck_collaboration_2018_I,planck_2018,planck_V_power_spectra_likelihoods}. This difference is usually quantified with the combined $S_8$ parameter, which is defined as $S_8 = \sigma_8\sqrt{\Omega_{\rm m}/0.3}$, with $\Omega_{\rm m}$ being the total matter budget of the Universe. If a fraction of the DM were allowed to decay, the clustering signal at low redshift would be modified, which might provide a solution to the $S_8$ difference in principle, as was pointed out by \citet[E15]{Enqvist_2015}, \cite{Berezhiani_2015_s8_1bddm}, \cite{chudaykin_2016_s8_1bddm} and \citet{Archidiacono_2019_s8_1bddm}. However, other authors have questioned these conclusions, showing that an agreement of the clustering amplitude between WL and the CMB cannot be easily achieved (\citealt[S22]{Simon_2022}; \citealt{mccarthy_s8_1bddm}).

Independent of the $S_8$ difference, several works have focused on providing forecasts and constraints for the one-body decaying dark matter (DDM) model using a variety of data from Milky Way satellite counts \citep{2022ApJ...932..128M}, WL shear observations (\eXV; \citealt[E20]{enqvist_2020}), and CMB data (\sXXII). Most authors have focused on the assumption that all DM is unstable, while models with decaying sub-species as part of a more complicated DM sector were investigated only little (\citealt{Poulin_2016}; \sXXII).

In the present paper, we study the effect of a one-body DDM fluid on the temperature and polarization spectra from \texttt{Planck} and on the WL band power spectrum from the latest \texttt{KiDS} data release. We use the Boltzmann solver {\tt Class} \citep{Blas_2011_CLASS,Julien_Lesgourgues_2011_noncold_relics} together with the nonlinear prescription of \cite{Hubert_2021} to model the effects of DM decay on the high- and low-redshift Universe. Our goal is on one hand to re-investigate the effect of one-body decay on the $S_8$ difference, and on the other hand, to provide new constraints on the half-life time of DDM and on the fraction of decaying to total DM.

The paper is structured in the following way: In Sec.~\ref{sec:1bDDM} we review  the theoretical aspects of the one-body DDM model. Sec.~\ref{sec:theory} and \ref{sec:inference} are dedicated to the presentation of our modelling pipeline, including the specifics of the Bayesian inference or Markov chain Monte Carlo (MCMC) process. In Sec.~\ref{sec:results} we present our results, before we conclude in Sec.~\ref{sec:conclusion}.
We benchmark our $\Lambda$CDM pipelines in Appendix~\ref{app:lcdm_benchmark} and provide more details about our MCMC analyses in Appendix~\ref{app:mcmc_details}.

\begin{figure}
    \centering
    \includegraphics{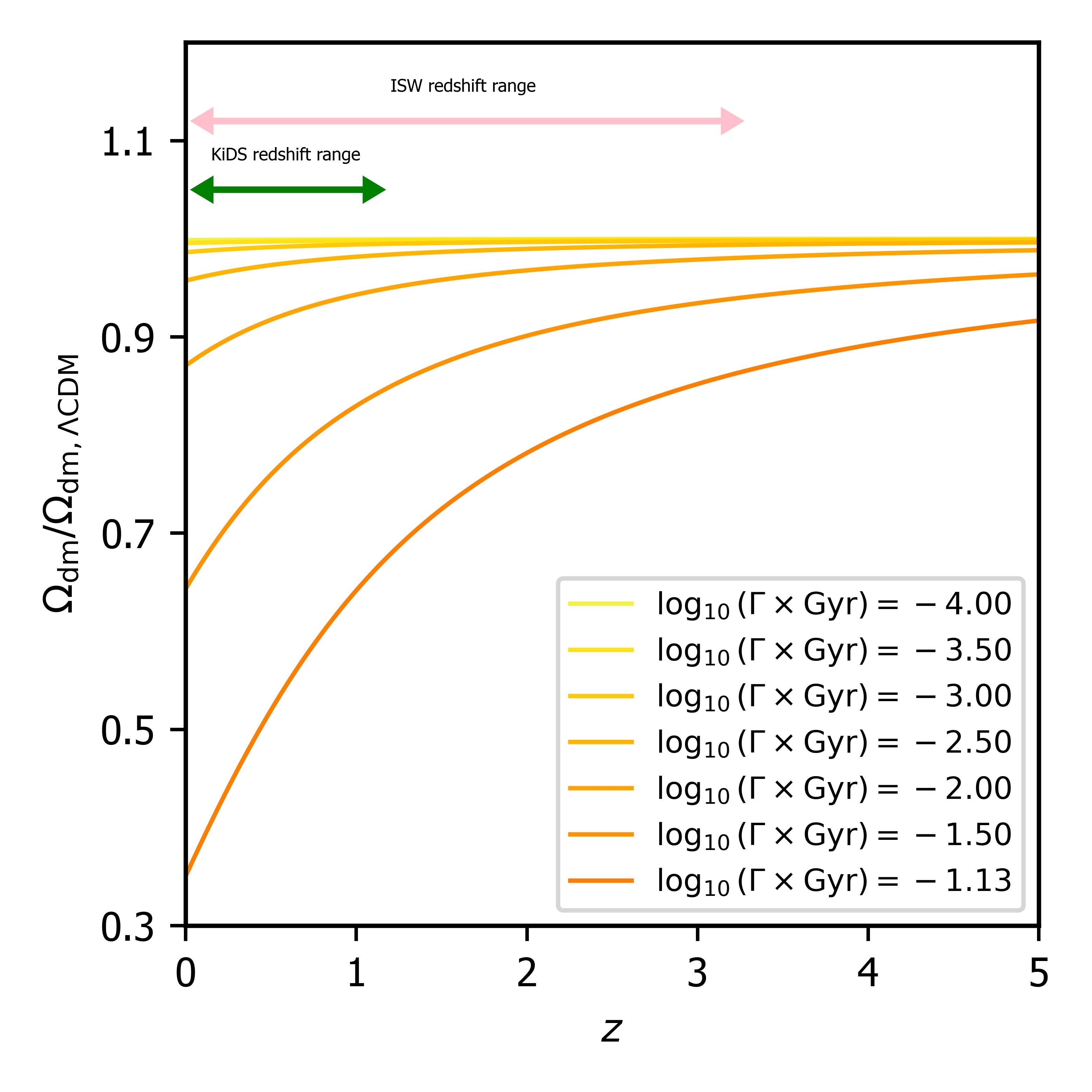}
    \caption{Redshift evolution of the DM abundance for different DM decay rates ($\Omega_\mathrm{dm}$) compared to the corresponding $\Lambda$CDM model ($\Omega_\mathrm{dm,\Lambda{CDM}}$). We assume that all DM is unstable ($f=1$). The green and pink arrows indicate the sensitivity ranges of the WL data from {\tt KiDS} and the ISW effect from {\tt Planck}.}
    \label{fig:Omega_z}
\end{figure}

\section{Decaying dark matter model}
\label{sec:1bDDM}

The DDM consists of a minimal extension of the standard $\Lambda$CDM model, where DM particles, instead of being stable, decay into massless relativistic particles propagating at the speed of light. A phenomenological description of this model includes two parameters (in addition to those describing the $\Lambda$CDM model), namely the decay rate of the DM particles $\Gamma$ and the fraction $f$ of decaying to total DM budget. As a result, the matter is transformed into radiation  affecting the background evolution of the Universe, that is,
\begin{eqnarray}
    &\label{eq:rho_DCDM}&{\rho}^{\prime}_{\rm dcdm} + 3\mathcal{H}\rho_{\rm dcdm} = -a\Gamma\rho_{\rm dcdm},\\
    &\label{eq:rho_DR}&{\rho}^{\prime}_{\rm dr} + 4\mathcal{H}\rho_{\rm dr} =a \Gamma \rho_{\rm dcdm},
\end{eqnarray}
where derivatives are expressed with respect to conformal time, $\mathcal{H}$ is a conformal Hubble parameter, and ${\rho}_{\rm dcdm}$ and ${\rho}_{\rm dr}$ are background densities of decaying cold DM and dark radiation, respectively (see e.g. \citet{Hubert_2021} for more details about the DM decay process). When only a fraction ($f$) of the total DM is allowed to decay, we define
\begin{eqnarray}
    & &f=\Omega_{\rm ddm,ini}/\Omega_{\rm dm,ini},\hspace{0.5cm} \Omega_{\rm dm,ini} = \Omega_{\rm ddm,ini} + \Omega_{\rm cdm,ini}
,\end{eqnarray}
where $\Omega_{\rm ddm,ini}$, $\Omega_{\rm cdm,ini}$, and $\Omega_{\rm dm,ini}$ are the decaying, stable, and total DM abundances at a time $t \ll \tau = \Gamma^{-1}$, that is, before the start of the decay process.

The background evolution of the Universe was modified as described in  Eqs.~\eqref{eq:rho_DCDM} and \eqref{eq:rho_DR}. In particular, the source terms whose amplitudes are set by  the decay rate $\Gamma$ cause a decrease in the DM and an increase in radiation abundance. In Fig.~\ref{fig:Omega_z} we show the evolution of the DM abundance between redshift 0 and 5 (solid lines). As expected, the DM abundance decreases towards low redshifts, whereas the amplitude of the effect depends on the decay rate ($\Gamma$). We also indicate the redshift range of the WL data from {\tt KiDS} as well as the range of late-time integrated Sachs-Wolfe (ISW) effect as measured by {\tt Planck} \citep[see e.g. ][]{Nishizawa_ISW}. Both observables overlap with the regime in which the effects of DDM are most prominent, making them promising probes to constrain DM decays.

\begin{figure*}
    \centering
    \includegraphics{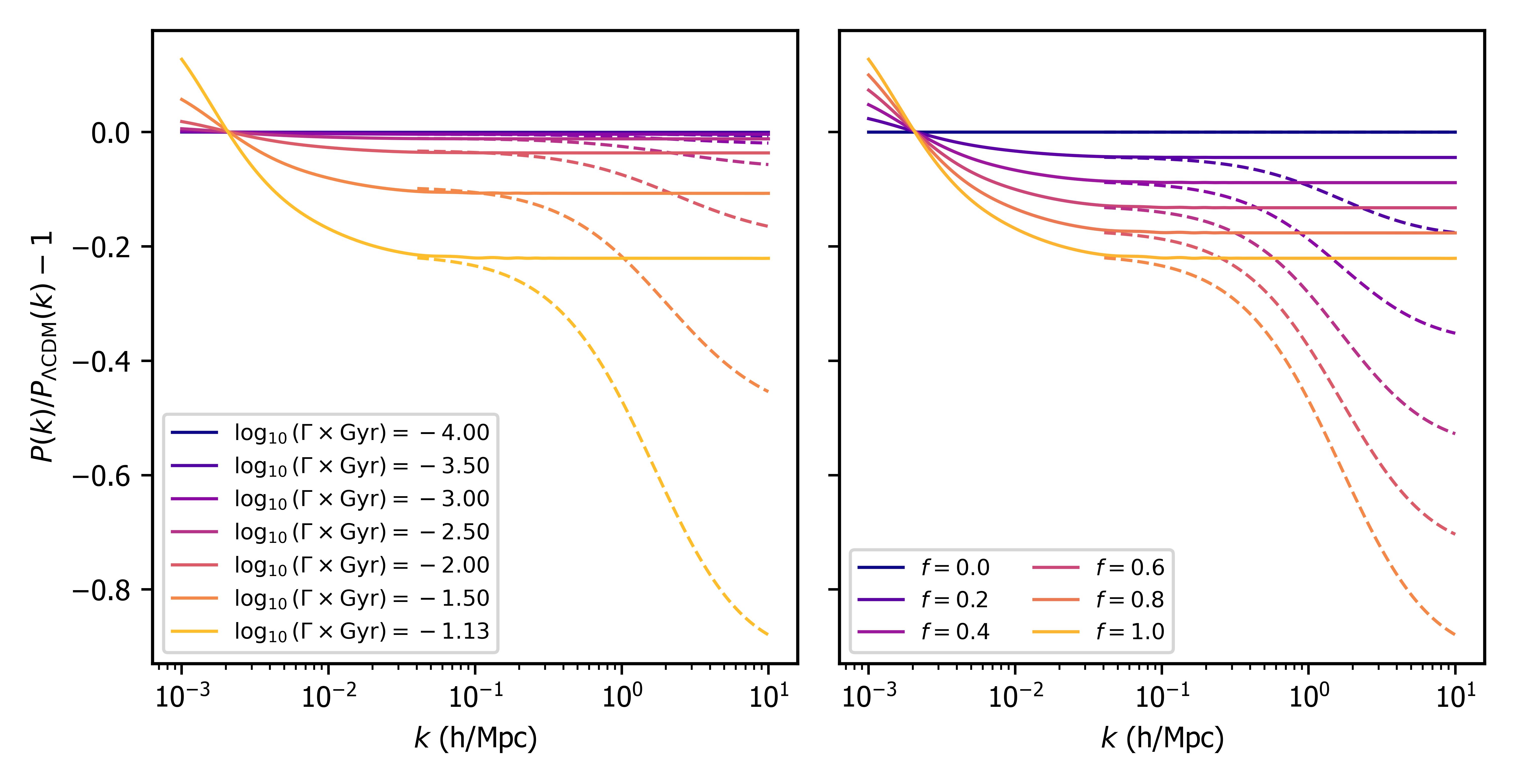}
    \caption{Suppression of the linear (solid) and nonlinear (dashed) matter power spectrum due to one-body DDM. The left panel shows the effects of varying the decay rate $\Gamma$ (in Gyr$^{-1}$) for a models in which all DM is allowed to decay ($f=1.0$). In the right panel, the fraction of decaying to total DM is varied, and the decay rate is fixed to $1/\Gamma = 13.5$~Gyr. }
    \label{fig:1bddm_param_effects}
\end{figure*}

The decay process affects not only the background evolution of the Universe, but also the process of structure formation.  Since the scale factor $a$ evolves at a somewhat slower rate (compared to $\Lambda$CDM), the Universe is less evolved, and the clustering process is therefore delayed. We are therefore left with suppression of power at small scales at a given redshift (see Fig.~\ref{fig:1bddm_param_effects}, described in the next section). This suppression  becomes more pronounced with high $\Gamma$ and with large $f$. Scenarios with $\Gamma \rightarrow 0$ and $f \rightarrow 0$ correspond to the $\Lambda$CDM model.

\section{Modelling pipeline}
\label{sec:theory}
In this section, we provide details of our  modelling pipeline for both the CMB and the WL observables. We specifically focus on nonlinear clustering, including the effects from DM decay, and we discuss our implementations of baryonic feedback and intrinsic alignment.

\subsection{Cosmic microwave background modelling}
\label{sec:cmb_modelling}

Although originating from the early Universe, the CMB temperature fluctuations provide strong constraints on reduced models, even for half-life times of the order of (or longer than) a Hubble time. The reason for this behaviour is the late-time ISW effect, which causes a modification of the large-scale CMB modes due to the gravitational redshifting of the CMB photons that pass through evolving potential wells. Following \cite{Nishizawa_ISW}, we can write the ISW spectrum as
\begin{eqnarray}
    &\label{eq:ISW_Cls}&
    C_l^{\rm ISW} = \frac{18}{\pi^2}\Omega_{\rm m}^2 H_0^4 \int {\rm d}k P(k)\left[\int {\rm d}r \mathcal{H}D(f_{\rm lin}-1)j_l(kr)\right]^2,
\end{eqnarray}
where $D$ is the linear growth factor, and $f_{\rm lin} = \frac{{\rm d}\ln D}{{\rm d}\ln a}$ is the velocity growth rate. The cosmological dependence of the ISW effect is governed by the term $\mathcal{H}D(f_{\rm lin}-1)$ and by the linear power spectrum. As discussed in \cite{Nishizawa_ISW}, the late-time ISW kernel starts to become important at $z\sim3-4$ steadily increasing towards $z\rightarrow 0$ when the Universe becomes dominated by dark energy.  In Fig~\ref{fig:Omega_z} we indicate the redshift range in which the CMB signal becomes sensitive to the ISW effect with a pink arrow.

To model the CMB including the ISW effect, we relied on the publicly available Boltzmann code {\tt Class}\footnote{\url{https://github.com/lesgourg/class_public}}, which comes with the option to include DM that decays into dark radiation. We modelled high-$\ell$ \textit{TT}, \textit{TE,} and \textit{EE} power spectra of the \textit{Planck 2018} CMB data using the lightweight version of the Planck \texttt{plik} likelihood, called \texttt{plik\_lite} \citep{planck_2020_XX_plik_lite,planck_V_power_spectra_likelihoods}, and mimicked \texttt{SimAll}  {(EE for $2\leq \ell <30$)} and \texttt{Commander}  {(TT for $2\leq \ell <30$)} likelihoods with the prior imposed on the optical depth parameter $\tau_{\rm reio}$ based on Eq.~4 of \citet[P20b]{planck_2018}.  {Even though this approximation was obtained for the $\Lambda$CDM scenario, the cosmological parameters recovered from CMB after one-body decay is included are very close to $\Lambda$CDM; see Tab.~\ref{tab:lcdm_results} and \ref{tab:dddm_results}. This allows for this approximation. Furthermore, \cite{Abellan_2021} compared the results of {\tt plik\_lite} and the full {\tt Plik} likelihoods for the more general scenario of two-body decays (which includes our model as a limiting case) and reported that the retrieved parameters agreed well.} The model parameters along with their prior ranges are listed in Tab.~\ref{tab:mcmc_parameters}. To evaluate the likelihood, we used all  215 data points for TT ($30\leq\ell\leq 2508$) and 199 data points for TE and EE ($30 \leq \ell \leq 1996$). We tested our inference pipeline for the  $\Lambda$CDM model and obtain an agreement of $\sim 0.1\sigma$ compared to the findings of the Planck Collaboration (see Appendix~\ref{app:lcdm_benchmark} and Fig.~\ref{fig:lcdm_planck_benchmark} for more details).

\subsection{Weak-lensing modelling}
To model WL cosmic shear observables, we followed the approach of \citet[Sch22]{schneider_2022_ksz}, with some changes as specified below. Most notably, we used the {\tt Pycosmo} package \citep{Refregier_2017pycosmo,pycosmo} combined with {\tt Class} to calculate the WL shear power spectra. For the nonlinear power spectrum, we relied on the revised halo model of \citet{Takahashi_2012_revised_halofit}. We included massive neutrinos with a fixed mass of 0.06 eV following the recipe from \planck. 
For the intrinsic alignment component, we used the nonlinear alignment model (NLA) introduced by \cite{bridle_2007} and described in \cite{hildebrandt_2016_kids_450}.  

In the following, we describe some other aspects of the modelling pipeline. We specifically focus on the implementation of DM decay, the handling of baryonic effects, and the connection to the band power data from {\tt KiDS}.

\subsubsection{Decaying dark matter}
\label{subsec:modelling_ddm}
To include the effects of one-body decay on the nonlinear matter power spectrum, we used the fitting function of \cite{Hubert_2021}, which corresponds to a modified version of the fit from \eXV. The function is defined by the ratio $P_{\rm DDM}(k,z)/P_{\rm \Lambda CDM}(k,z) = 1 - \varepsilon_{\rm nonlin}(k,z)$, where
\begin{eqnarray}&\label{fit}&
\frac{\varepsilon_{\rm nonlin}(k,z)}{\varepsilon_{\rm lin}} = \frac{1+a(k/\text{Mpc}^{-1})^p}{1+b(k/\text{Mpc}^{-1})^q}f,
\end{eqnarray}
with the factors $a$, $b$, $p$, and $q$ given by
\begin{eqnarray}
& &a(\tau,z) = 0.7208 + 2.027\left(\frac{\text{Gyr}}{\tau}\right) + 3.031\left(\frac{1}{1+1.1z}\right) - 0.18,\nonumber\\
& &b(\tau,z) = 0.0120 + 2.786\left(\frac{\text{Gyr}}{\tau}\right) + 0.6699\left(\frac{1}{1+1.1z}\right) - 0.09,\nonumber\\
& &p(\tau,z) = 1.045 + 1.225\left(\frac{\text{Gyr}}{\tau}\right) + 0.2207\left(\frac{1}{1+1.1z}\right) - 0.099,\nonumber\\
& &q(\tau,z) = 0.992 + 1.735\left(\frac{\text{Gyr}}{\tau}\right) + 0.2154\left(\frac{1}{1+1.1z}\right) - 0.056.\nonumber
\end{eqnarray}
The remaining function $\varepsilon_{\rm lin}$ describes the redshift evolution of the suppression and is given by
\begin{eqnarray}&\label{eq:eps_lin}&
\varepsilon_{\rm lin}(\tau, z) = \alpha\left(\frac{\rm Gyr}{\tau}\right)^{\beta}\left(\frac{1}{(0.105z)+1}\right)^{\gamma},
\end{eqnarray}
where $\alpha$, $\beta$, $\gamma$ are functions of $\omega_b$, $h$, and $\omega_{\rm m} = \omega_{\rm b} + \omega_{\rm  dm}$, that is,
\begin{eqnarray}
\quad\quad\alpha  &=&  (5.323 - 1.4644u - 1.391v) + (-2.055 + 1.329u \nonumber\\ 
&&+0.8672v)w + (0.2682 - 0.3509u)w^2,\nonumber\\
\beta  &=&  0.9260 + (0.05735 - 0.02690v)w \nonumber\\&&+ (-0.01373 + 0.006713v)w^2,\nonumber\\
\gamma  &=&  (9.553 - 0.7860v) + (0.4884 + 0.1754v)w \nonumber\\ &&+ (-0.2512 + 0.07558v)w^2.\nonumber
\end{eqnarray}
We defined $u=\omega_b/0.02216$, $v=h/0.6776$ and $w = \omega_m/0.14116$. The fitting function is able to reproduce results from $N$~-~body simulations with an error smaller than 1\%\ up to $k=13$ h/Mpc \citep{Hubert_2021}. In order to calculate the DDM matter power spectrum at nonlinear scales, we multiplied the term $(1-\varepsilon_{\rm nonlin})$ with the $\Lambda$CDM power spectrum from the revised halofit model of \citet{Takahashi_2012_revised_halofit}.

In Fig.~\ref{fig:1bddm_param_effects} we illustrate the effect of DDM on the linear (solid lines) and nonlinear (dashed lines) matter power spectrum. Different colours correspond to different decay rates ($\Gamma$) for a fixed $f=1$ (left panel) and different fractions ($f$) for a half-life time $1/\Gamma=13.5$ Gyr (right panel). In general, DM decay leads to a suppression of power towards small scales. This effect is amplified by nonlinear clustering. The power suppression can be understood by the fact that the clustering in the DM model is delayed compared to $\Lambda$CDM, causing galaxy groups and clusters (which dominate the power spectrum signal) to form later.

\subsubsection{Baryonic feedback}
\label{sec:bayonic_feedback}

Baryonic feedback effects play an important role in the WL signal \citep[e.g.][]{chisari,vanDaalen:2019pst,arico_2021_bacco}. They lead to suppression of the matter power spectrum, which may be of similar shape to the suppression due to DDM \citep{Hubert_2021,Amon:2022azi}. In order to account for potential degeneracies between the DM and the baryonic sector, it is therefore particularly important to model baryonic effects in the DDM cases. 

We used the emulator \texttt{BCemu} \citep{Giri_2021_bcemu}, which includes the effects of baryonic feedback on the matter power spectrum. {\tt BCemu} is based on the \textit{baryonification} model described in \cite{Schneider_2015_bfc} and \cite{Schneider_2018_bfc}. It has seven free model parameters describing the specifics of the gas and stellar distributions around haloes, as well as one cosmological parameter that is the baryon ($f_b=\Omega_b/\Omega_m$). We fixed four of the seven parameters and only varied the gas parameters $\log_{10} M_{\rm c}$, and $\theta_{\rm ej}$, as well as the stellar parameter $\eta_\delta$. Furthermore, the baryon fraction $f_b$ was varied in accordance with the cosmological parameters. This three-parameter model has been shown in \citet{Giri_2021_bcemu} to match the power spectra from hydrodynamical simulations at the percent level for $k\lesssim 12.5$~h/Mpc.

\subsubsection{Cosmic shear angular power spectrum with \textit{KiDS-1000}}
\label{subsec:cosmic_shear}

The latest catalogue released by the Kilo-Degree Survey (\textit{KiDS-1000}) contains shear information of over 20 million galaxies distributed inside five tomographic bins between $z\sim0.1$ and $z\sim 1.2$ \citep{Kuijken_2019_kids}. We used the band power spectrum published in \aXXI~using the auto and cross spectra of all five tomographic bins. The corresponding covariance matrix is from \citet{joachimi_2020}.

To model the cosmic shear power spectrum components from gravitational lensing (G) and the intrinsic alignment of galaxies (I), we used the modified Limber approximation \citep{loverde_ext_limber_2008,kilbinger_ext_limber_2017}, that is,

\begin{eqnarray}
    & &C_{\rm AB}^{(ij)}(\ell) = \int_0^{\chi_{\rm h}} {\rm d}\chi \frac{W^{(i)}_{\rm A}(\chi)W^{(j)}_{\rm B}(\chi)}{f^2_{\rm K}(\chi)}P_{\rm m}^{\rm nonlin}\left(\frac{\ell+1/2}{f_{\rm K}(\chi)},z(\chi) \right),
    \label{eq:C_AB}
\end{eqnarray}
where ${\rm A,B}\in \{\rm G,I\}$. $\chi$ is the comoving radial distance, and $f_{\rm K}(\chi)$ is the comoving angular diameter distance. 
The window functions of the gravitational and intrinsic alignment components are given by
\begin{eqnarray}
    &\label{eq:W_G}&
    W^{(i)}_{\rm G}(\chi) = \frac{3H_0^2\Omega_{\rm m}}{2c^2}\frac{f_{\rm K}(\chi)}{a(\chi)}\int_\chi^{\chi_{\rm h}} {\rm d}\chi' n_{\rm S}^{(i)}(\chi') \frac{f_{\rm K}(\chi' - \chi)}{f_{\rm K}(\chi')},\\
    &\label{eq:W_I}&
    W^{(i)}_{\rm I}(\chi) = -A_{\rm IA} \left(\frac{1+z(\chi)}{1+z_{\rm pivot}}\right)^{\eta_{\rm IA}} \frac{C_1 \rho_{\rm cr}\Omega_{\rm m}}{D(a(\chi))} n_{\rm S}^{(i)}(\chi),
\end{eqnarray}
where $D(a)$ is the linear growth factor, and the $n_s^{(i)}$ terms correspond to the redshift distribution of source galaxies for each tomographic bin ($i$). The term $C_1 \rho_{\rm cr}$ was fixed to $0.0139,$ and $z_{\rm pivot}$ was set to 0.3 \citep[see][]{joachimi_2011}.

From the angular shear power shown in Eq.~(\ref{eq:C_AB}), we calculated the band power spectrum following \cite{joachimi_2020}. We refer to \schXXII~for more details about this procedure.  {The prescription for cosmic shear modelling above does not strictly rely on $\Lambda$CDM. In our case, all relevant changes to the modelling enter via modifications of the nonlinear matter power spectrum.}

\section{Model inference}
\label{sec:inference}

\begin{table}
    \tiny
    \renewcommand{\arraystretch}{1.3}
    \centering
    \begin{tabular}{l l c c}
    \hline
    \hline
    \rule{0em}{-0.5em}\\
    Parameter name & Acronym     & prior  & range  \\
    \hline
(Initial) cold DM abundacne & $\omega_{\rm c}$          &       flat    &       [0.051, 0.255]\\ 
Baryon abundance & $\omega_{\rm b}$             &       flat    &       [0.019, 0.026]\\ 
Scalar amplitude & $\ln (10^{10} A_{\rm s})$    &               flat    &       [1.0, 5.0]\\ 
Hubble constant &$h_0$  &       flat    &       [0.6, 0.8]\\ 
Spectral index &$n_{\rm s}$     &       flat    &       [0.9, 1.03]\\  
Optical depth &$\tau_{\rm reio}$        &       normal  &       $\mathcal{N}(0.0506,0.0086)$\\ 
\hline
Intrinsic alignment amplitude &$A_{\rm IA}$             &               flat    &       [0.0, 2.0]\\ 
Planck calibration parameter &$A_{\rm planck}$  &               normal  &       $\mathcal{N}(1.0,0.0025)$\\ 
\hline
First gas parameter (\texttt{BCemu}) &$\log_{10} M_{\rm c}$             &               flat    &       [11.0, 15.0]\\ 
Second gas parameter (\texttt{BCemu}) &$\theta_{\rm ej}$        &               flat    &       [2.0, 8.0]\\ 
Stellar parameter (\texttt{BCemu}) &$\eta_\delta$       &               flat    &       [0.05, 0.40]\\ 
\hline
Decay rate & $\log_{10} \Gamma$ &               flat    &       [$-4.00$, $-1.13$]\\  
Fraction of DDM &$ f$   &               flat    &       [0.0,1.0]\\ 
    \hline
    \end{tabular}
    \caption{Parameters and choices of priors employed in our MCMC analysis. \textit{Flat} denotes a uniform prior within boundaries specified in the last columns. For the Gaussian prior (\textit{normal}), we state the mean and standard deviation.}
    \label{tab:mcmc_parameters}
\end{table}

\begin{figure*}[!th]
     \centering
     \begin{subfigure}[b]{0.47\textwidth}
         \centering
         \includegraphics[width=\textwidth]{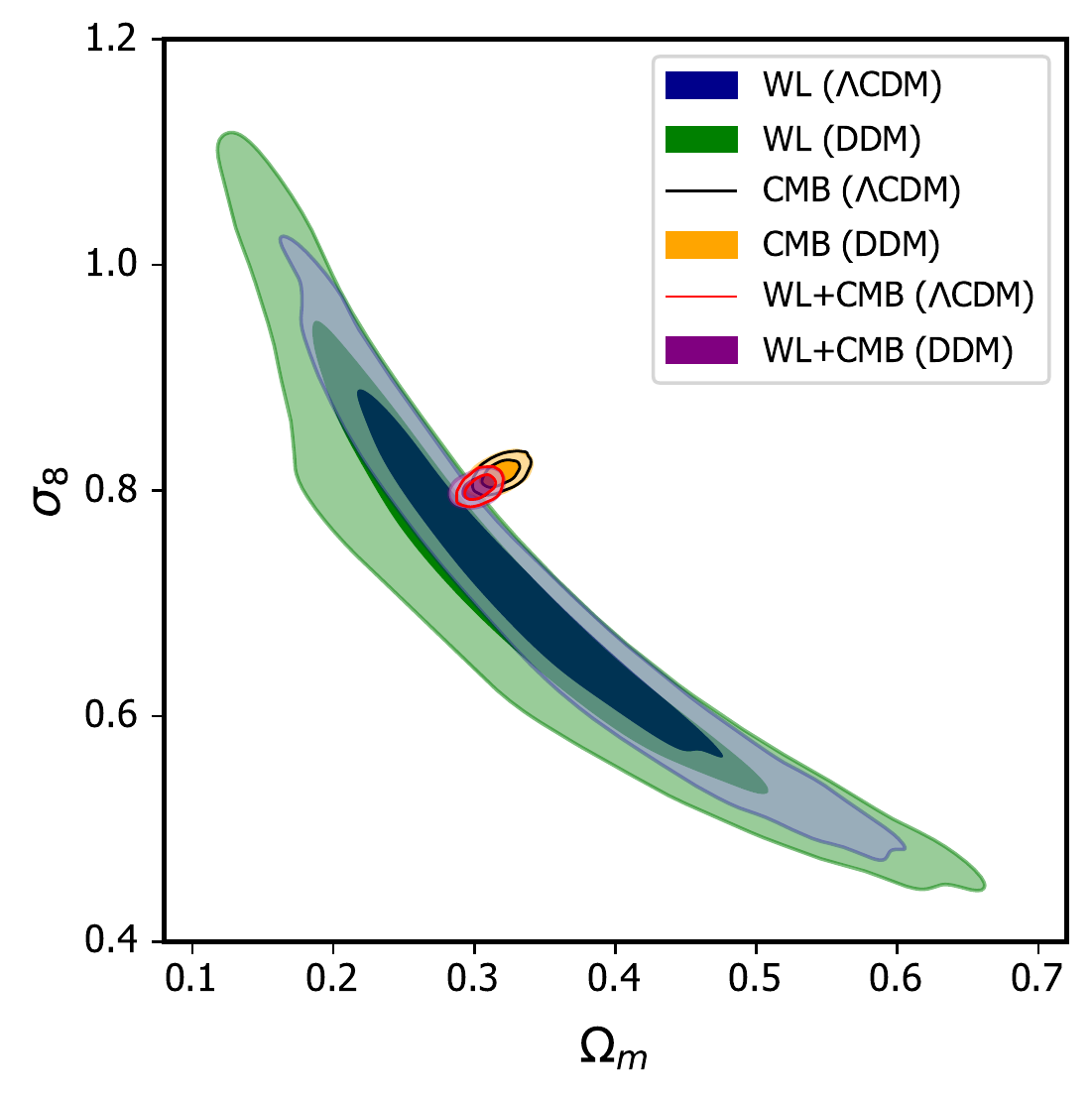}
         \label{fig:S8_a}
     \end{subfigure}
     \begin{subfigure}[b]{0.476\textwidth}
         \centering
         \includegraphics[width=\textwidth]{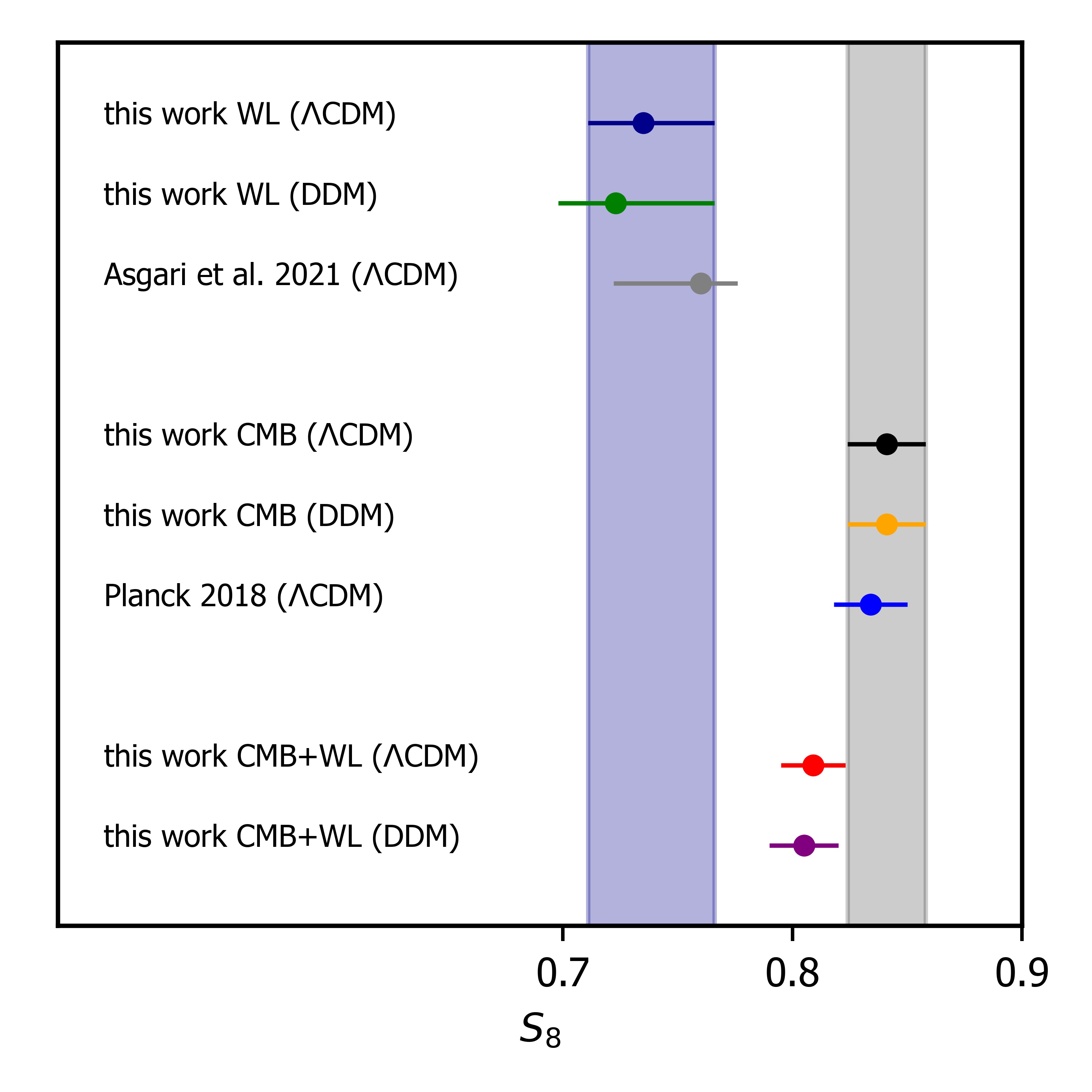}
         \label{fig:S8_b}
     \end{subfigure}
        \caption{Implications of two-body decays on clustering amplitude $\sigma_8$ and on $S_8$ parameter from our analysis of CMB and WL data. {\it Left:} Two-dimensional posterior contours (68\% and 95\% confidence levels) of the $\Omega_{\rm m} - \sigma_8$ plane from our MCMC runs for the $\Lambda$CDM and DDM scenario. Results from \textit{KiDS-1000} are shown in blue and green, and results from \textit{Planck 2018} are shown in black and yellow. The combined CMB+WL analysis is shown in red and purple. {\it Right:} One-dimensional constraints of the $S_8$ parameter for the same models. The original results from the {\it KiDS-1000} \citep[A21]{kids_1000} and the {\it Planck 2018} \citep[P20b]{planck_2018} analyses are added in grey and blue for comparison.}
        \label{fig:S8}
\end{figure*}

We used the \texttt{emcee} package \citep{emcee} with the stretch move ensemble method in our MCMC analyses. For the WL and the CMB setup, we assumed multivariate Gaussian likelihoods. The convergence of the chains was checked with the \textit{Gelman-Rubin} criterion assuming $R_c<1.1$ \citep{Gelman_Rubin_1992}. In the case of the CMB analysis, we used the covariance matrix provided alongside the \texttt{Plik\_lite} likelihood. For the WL analysis, we relied on the band power covariance matrix published by the KiDS collaboration \citep{joachimi_2020}.

In Table~\ref{tab:mcmc_parameters} we provide a summary of all model parameters, including information about their priors. For the CMB analysis and the WL analysis, we sampled over 9 and 12 parameters, respectively. The combined chains contain 13 free parameters.
We used flat priors for all cosmological parameters except for the optical depth $\tau_{\rm reio}$, for which we assumed a Gaussian prior with a mean $\tau_{\rm reio}=0.0506$ and standard deviation $\sigma_\tau = 0.0086,$ as explained in  Sec.~\ref{sec:cmb_modelling}. For cold DM abundance $\omega_{\rm c}$ and primordial power spectrum amplitude $A_{\rm s}$, we used a prior wide enough to be uninformative. In the DDM scenario, $\omega_{\rm c}$ stands for the initial cold DM abundance. In terms of \texttt{CLASS} input variables, we set \texttt{omega\_cdm} = $(1-f) \omega_{\rm c}$ and \texttt{omega\_ini\_dcdm} = $f\omega_{\rm c}$.  In the case of parameters for which WL alone is not sensitive enough ($\omega_{\rm b}, h_0, n_{\rm s}$, $\log_{10} M_c$, $\theta_{\rm ej}$, and $\eta_{\delta}$), we defined wide prior ranges following the analyses in \aXXI~and \schXXII. Regarding the baryonic parameters, the prior ranges are limited by the range of the emulator. They comfortably include all results from hydrodynamical simulations, however \citep{schneider_2020_baryonic_effects2,schneider_2020_baryonic_effects_1,Giri_2021_bcemu}. For the Planck absolute calibration $A_{\rm planck}$ we followed the suggestion of the Planck Collaboration\footnote{\url{https://wiki.cosmos.esa.int/planck-legacy-archive/index.php/CMB_spectrum_\%26_Likelihood_Code}}  and choose Gaussian prior $\mathcal{N}(1.0,0.0025)$. The adopted intrinsic alignment model (NLA) assumes two free parameters $A_{\rm IA}$ and $\eta_{\rm IA}$ entering via Eq.~\eqref{eq:W_I} of Sec.~\ref{subsec:cosmic_shear}. Following \aXXI, for example, we set $\eta_{\rm IA}=0$, and kept only $A_{\rm IA}$ as a free parameter.

We ran six chains in total, three assuming a $\Lambda$CDM cosmology, and three including the possibility of DM decay. The three runs refer to the CMB alone, the WL alone, and the combined setup. The main results from these chains in terms of DM constraints and cosmology are shown in the next section. Further details are provided in Appendix~\ref{app:mcmc_details}, where we list the best-fit values and errors for all the parameters involved in the MCMC analysis.

\section{Results}
\label{sec:results}
The main goal of this paper is to constrain DM decays with Planck and \textit{KiDS-1000} data. However, before showing the obtained limits on the decay rate and the fraction of decaying to total DM, we discuss the effect of the DDM scenario on the $S_8$  difference. 

In the left panel of Fig.~\ref{fig:S8}, we show the posterior contours of the $\sigma_8$-$\Omega_m$ plane for our different data and modelling choices. For the case of $\Lambda$CDM, the results from KiDS and Planck are shown in black and blue, respectively. The best-fit values and 68\%\  errors of the combined $S_8$ parameter are given by
\begin{eqnarray}
    & &S_8 = 0.735^{+0.031}_{-0.024} \quad ({\rm WL}, \Lambda\rm CDM),\\
    & &S_8 = 0.841 \pm 0.017 \quad ({\rm CMB}, \Lambda\rm CDM),
\end{eqnarray}
corresponding to a difference of 3.0$\sigma$,  {which we obtained using the same conventional method as was used in \aXXI \, [see their eq.~(16)]}. These findings agree well with the original results from the KiDS (\aXXI) and Planck (\planck) collaborations, as shown in the right panel of Fig.~\ref{fig:S8} and Appendix~\ref{app:lcdm_benchmark}.

The posterior contours of the DDM case are shown in yellow and green for Planck and KiDS, respectively. They do not show any visible shift with respect to the $\Lambda$CDM case, except that the KiDS contours become broader, especially towards lower values of $\sigma_8$ and $\Omega_m$. We assume this to be the result of degeneracies between the baryonic and DDM parameters. Regarding the combined $S_8$ parameter, the best-fitting values and 68\%\  errors are given by
\begin{eqnarray}
    & &S_8 = 0.723^{+0.041}_{-0.027} \quad ({\rm WL}, \rm DDM),\\
    & &S_8 = 0.841 \pm 0.017 \quad ({\rm CMB}, \rm DDM),
\end{eqnarray}
yielding an $S_8$  difference of 2.7$\sigma$. This small decrease in the difference is not due to a better concordance of the $S_8$ values, but rather to a general increase in the error budget in the DDM case of the constraints derived from the WL data.

The above point can be further quantified by investigating the decrease in the minimum chi-squared ($\chi_{\rm min}^2$) from the standard $\Lambda$CDM to the DDM model. The change in the Akaike information criterion $\Delta {\rm AIC} = \Delta \chi^2_{\rm min} + 2(N_{\rm DDM} - N_{\rm \Lambda CDM})$, which compensates for the increase in the goodness of fit due to the increased parameter space, gives
\begin{eqnarray}
    & &\Delta {\rm AIC} =  {4.0} \quad \rm (WL),\\
    & &\Delta {\rm AIC} = 3.9 \quad \rm (CMB)
\end{eqnarray}
for the WL and the CMB case. In the definition of $\Delta \rm AIC$, $\Delta \chi^2_{\rm min}$ stands for the difference of $\chi^2_{\rm min}$ between DDM and $\Lambda$CDM and $N_{\rm DDM}$ ($N_{\rm \Lambda CDM}$) denotes the number of free parameters in the DDM ($\Lambda$CDM) model. Despite two more parameters, the decrease in $\Delta \chi^2_{\rm min}$ is not sufficient in the DDM case compared to $\Lambda$CDM (models for which the increased 
number of free parameters is compensated for by the better goodness-of-fit result in  $\Delta {\rm AIC}<0$). 

Although there is a remaining difference between the $\Lambda$CDM and DDM models, we ran combined chains for both scenarios. In Fig~\ref{fig:S8} they are given by the red and purple contours (left panel) and data points (right panel). As expected, the posteriors from the combined MCMC runs are located in between the two original contours. Based on these results, we can now quantify the general difference between the KiDS and Planck datasets. Following \citet{raveri_hu_qmap_2019}, we defined a difference in the maximum a posterior (MAP), which takes the full multi-dimensional posterior distribution into account,
\begin{eqnarray}
    & &Q_{\rm MAP} = \chi^2_{\rm min,comb} - \left(\chi^2_{\rm min,Planck} + \chi^2_{\rm min,KiDS}\right).
\end{eqnarray}
With this definition, the difference between the two probes can be expressed as $T = \sqrt{Q_{\rm MAP}}$. For the $\Lambda$CDM and the DDM scenarios, we obtain
\begin{eqnarray}
    & &T =  {3.4}\sigma  \quad (\Lambda\rm CDM),\\
    & &T = 3.4\sigma \quad (\rm DDM).
\end{eqnarray}
From the difference in MAP, we conclude that one-body decays do not lower the mutual difference between \textit{KiDS-1000} and \textit{Planck 2018} data. A summary of the minimum $\chi^2$~values as well as the scores of the AIC and MAP is provided in Tab.~\ref{tab:S8_summary}.

\begin{table}[]
    \renewcommand{\arraystretch}{1.3}
    \centering
    \begin{tabular}{l|c|c|c|c}
                        & KiDS   & Planck 2018 & Combined   & $\sqrt{ Q_{\rm MAP}}$  \\
        \hline
        \hline
        $\chi^2_{\rm min}$ ($\Lambda$CDM)    &  {158.7}  &  580.2     &  {750.5}      &  {3.4}$\sigma$ \\ 
        $\chi^2_{\rm min}$ (DDM)             & 158.7  &  580.1     & 750.2      & 3.4$\sigma$ \\ 
        \hline
        $\Delta \chi^2_{\rm min}$ &  {0.0}    &   {-0.1}       &   {-0.3}      &  \\
        $\Delta$ AIC    &  {4.0}    &  3.9       &  {3.7}      & 
    \end{tabular}
    \caption{Minimum $\chi^2$~~values from inference with KiDS and Planck data separately as well as from the combined run. In the last column, the difference between the two datasets as viewed by \textup{\textit{\textup{difference in maximum a posteriori}} }criterion. The last two rows show the difference of the minimum $\chi^2$ between the $\Lambda$CDM and DDM cases, subtracting the first from the second, and the balance of the goodness-of-fit improvement and the increase in the number of parameters using the Akaike information criterion.}
    \label{tab:S8_summary}
\end{table}

\begin{figure}[!t]
    \centering
    \includegraphics[width = \columnwidth]{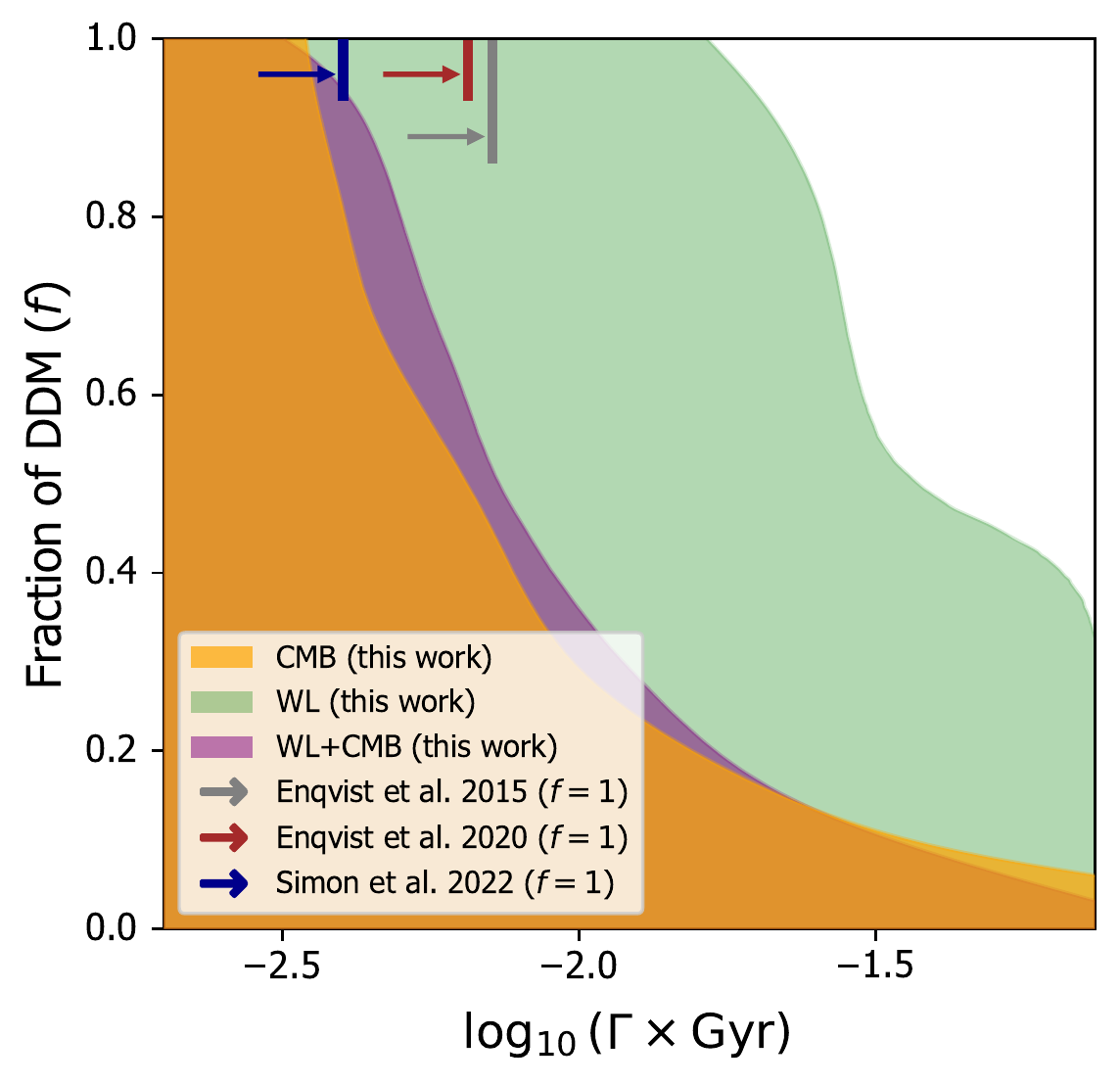}
    \caption{Constraints on the decay rate ($\Gamma$) and the decaying-to-total DM fraction ($f$) for the one-body DDM scenario. Weak lensing (WL) results from \textit{KiDS-1000} and CMB results from \textit{Planck 2018} are shown in green and yellow. The combined CMB + WL analysis is shown in purple. For the limiting case of $f=1$, we add other results from recent studies by  \cite[E15]{Enqvist_2015} (grey arrow), \cite[E20]{enqvist_2020} (red arrow), and \cite[S22]{Simon_2022} (blue arrow). All results are provided at the 95\% confidence level.}
    \label{fig:Gamma_f_compare}
\end{figure}

We now turn our attention towards the constraints on one-body decay obtained by the CMB, WL, and combined datasets used in this paper. The two-dimensional constraints for the DDM parameters $\Gamma$ and $f$ are illustrated in Fig.~\ref{fig:Gamma_f_compare}. All limits are  provided at the 95\% confidence levels. The contours exhibit the expected hyperbolic shape, excluding the regime in the top right corner of high decay rates and larger fractions of decaying to total DM. The results from Planck (yellow contours) show much stronger constraining power than those from the KiDS data. This means that the ISW effect is currently more sensitive to DM decay than WL. However, this is likely to change in the near future due to new WL observations from Euclid \citep{Hubert_2021}.

The combined CMB + WL constraints, shown as purple contours in Fig.~\ref{fig:Gamma_f_compare}, are comparable in strength to the CMB-only limits. The small differences between $f=0.2-0.9$ are most likely caused by the inherent differences between the KiDS and Planck datasets. A similar behaviour has been reported by \eXV. 

In Fig.~\ref{fig:Gamma_f_compare} we compare our results to several recent studies from the literature. Because these studies only provide constraints for the limiting case of $f=1$, they were added as arrows in the top part of the plot. We show findings from \eXV, \eXX, and \sXXII. \eXV~used Planck 2013 data combined with nine-year WMAP polarization measurements and WL data from CFHTLens \citep{CFHTLens}, reporting $\Gamma^{-1} \geq 140$~Gyr. \eXX~combined Planck 2015 CMB data with Planck 2015 SZ cluster counts and KiDS 450 WL observations obtaining $\Gamma^{-1} \geq 154$~Gyr. \sXXII~combined CMB data from \textit{Planck 2018} with the Pantheon dataset and BAO from BOSS, 6dFGS, and SDSS DR7. They reported $\Gamma^{-1} \geq 260.4$~Gyr. 

For our limiting case of $f=1$, we obtain a half-life time of $\Gamma^{-1}\geq 288$~Gyr. This limit is slightly stronger than that of \sXXII~and is significantly stronger than those of \eXV~and \eXX. Compared to \eXV~and \eXX, we used a more recent dataset which is probably responsible  for strengthening the constraints. 
Compared to \sXXII, the differences are much smaller, which is expected because both studies used \textit{Planck 2018} data for the analysis.
For the case of high decay rates and small decaying to total DM fractions, we obtain limits of $f<0.34$, $f<0.07$, and $f<0.03$ for  the KiDS, Planck, and the combined analysis at the 95\% confidence level.

\section{Conclusions}
\label{sec:conclusion}
We have investigated the one-body DDM scenario and its effects on structure formation in the light of CMB TTTEEE data from \textit{Planck 2018} and the cosmic shear angular power spectra from the \textit{KiDS-1000} data release. The free parameters of the DDM model are the decay rate ($\Gamma$) and the decaying to total DM fraction ($f$). 

We obtained new constraints on $\Gamma$ and $f$ from the CMB, from WL, and from the combined CMB + WL analysis. In agreement with previous results, we find that the CMB constraints are stronger than those from WL alone. This apparently surprising result is due to the ISW effect, which provides strong constraints on the late-time background evolution of the Universe.

For the limiting case of $f=1$, we obtain $\Gamma^{-1} \geq 288$~Gyr, which is stronger than previous constraints from \eXV~and \eXX~and similar to the findings of \sXXII. For high decay rates ($\Gamma\sim H_0$), on the other hand, we find a limit on the decaying to total DM fraction of $f<0.03$, which is based on the combination of CMB and WL data. The CMB alone provides weaker constraints of $f<0.07$. 

Along with the derivation of new constraints on the one-body DDM scenario, we also investigated the effect of decay on the $S_8$  difference reported for example by the KiDS collaboration \citep{heymans_2021_CFHTLenS}. At face value, we find a slight reduction of the difference from 3.0$\sigma$ to 2.7$\sigma$ from a $\Lambda$CDM to a DDM model. We showed, however, that this reduction is entirely caused by the increase in free parameters. Our maximum a posteriori probability analysis (MAP) yields no improvement from a $\Lambda$CDM to a DDM scenario.

We conclude that there is currently no evidence for a DM sector featuring one-body decay from matter to radiation. For most of the parameter space, current WL observations are not constraining enough to compete with the stringent limits obtained from the CMB radiation via the integrated Sachs-Wolfe effect. In the near future, however, results from stage-IV lensing surveys such as Euclid are expected to probe currently untested DDM scenarios.

\begin{acknowledgements}
This work is supported by the Swiss National Science Foundation under the grant number PCEFP2\_181157. 
Nordita is supported in part by NordForsk.
\end{acknowledgements}

\bibliography{bucko_1bDDM.bib}
\bibliographystyle{aa}

\appendix

\section{$\Lambda$CDM benchmark}
\label{app:lcdm_benchmark}

\begin{figure}
    \centering
    \includegraphics[width = \columnwidth]{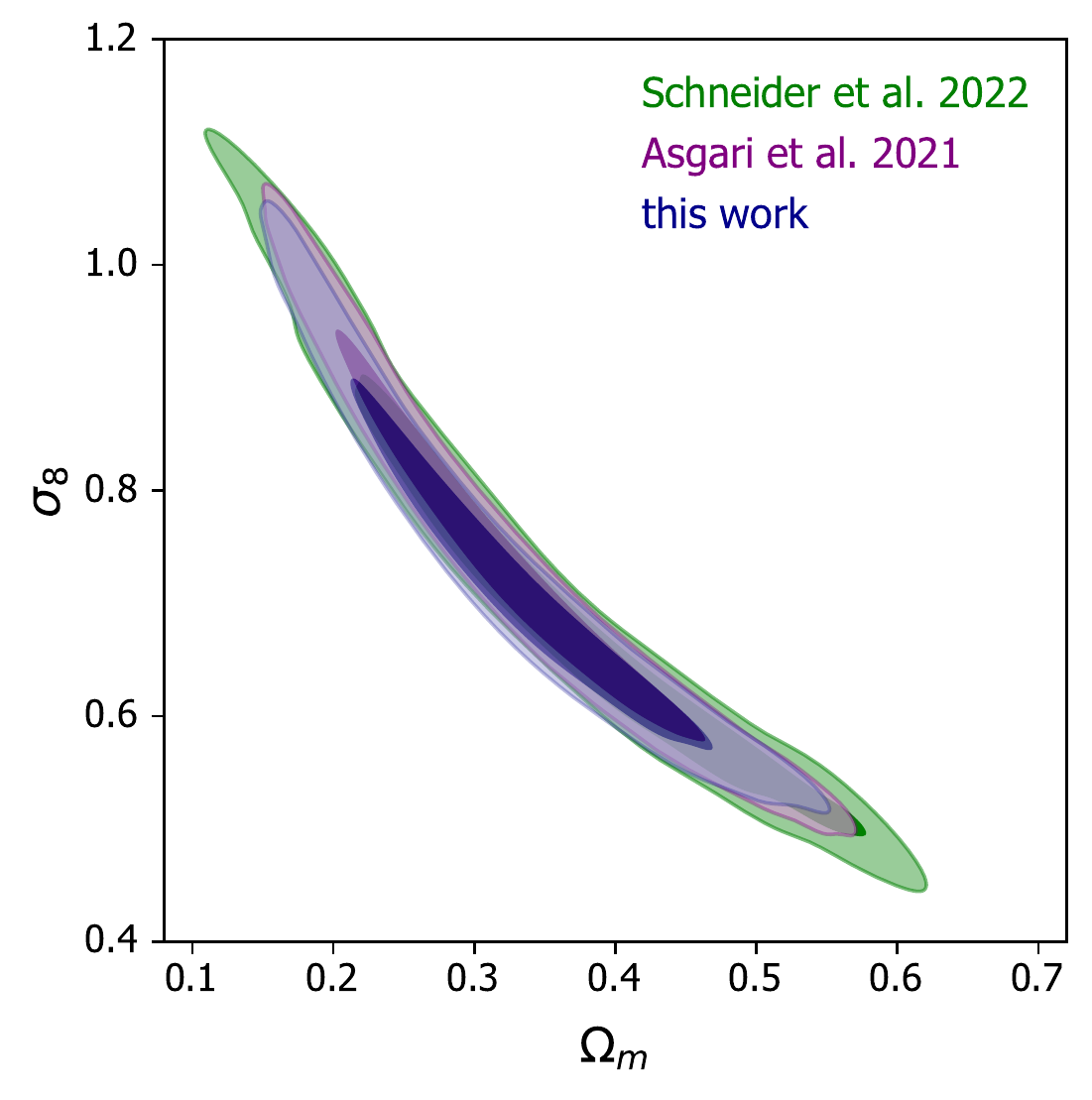}
    \caption{Comparison of $\Lambda$CDM posterior contours in the $\Omega_{\rm m}-\sigma_8$~-~plane obtained in this work with two recent studies that modelled the cosmic shear band power signal.}
    \label{fig:lcdm_benchmark}
\end{figure}

We compared the results of our pipeline in the case of $\Lambda$CDM to the original results of KiDS collaboration \aXXI~and to \schXXII~using a similar approach of band power modelling. For brevity, we only present the $\Omega_{\rm m} - \sigma_8$ contours shown in Fig.~\ref{fig:lcdm_benchmark}. The difference in $1$- and $2\sigma$ contours is marginal compared to the original \textit{KiDS-1000} results. The most significant differences (to our best knowledge) arise from the choice of baryonic prescription; the \textit{KiDS-1000} pipeline uses the one-parametric \texttt{HMCODE} baryonic feedback model \citep{mead_hmcode_2015}, while this work adopted a four-parametric (three baryonic parameters plus the baryon-to-matter ratio)
version of \texttt{BCemu} \citep{Giri_2021_bcemu}.  Compared with \schXXII, the 68\% and 95\% confidence intervals are slightly more extended. The modelling pipelines, which otherwise are very similar, employ a different number of baryonic parameters, specifically, eight (seven baryonic parameters plus the baryon-to-matter ratio) in the case of \schXXII~  and four in this work. This  results in broader posterior contours in the case of \schXXII.  {Quantitatively, our results agree better with those of \aXXI\, at $0.2\sigma$ for $\sigma_8$ and at the level of $0.1\sigma$ for $\Omega_{\rm m}$. }

\begin{figure}
    \centering
    \includegraphics[width = \columnwidth]{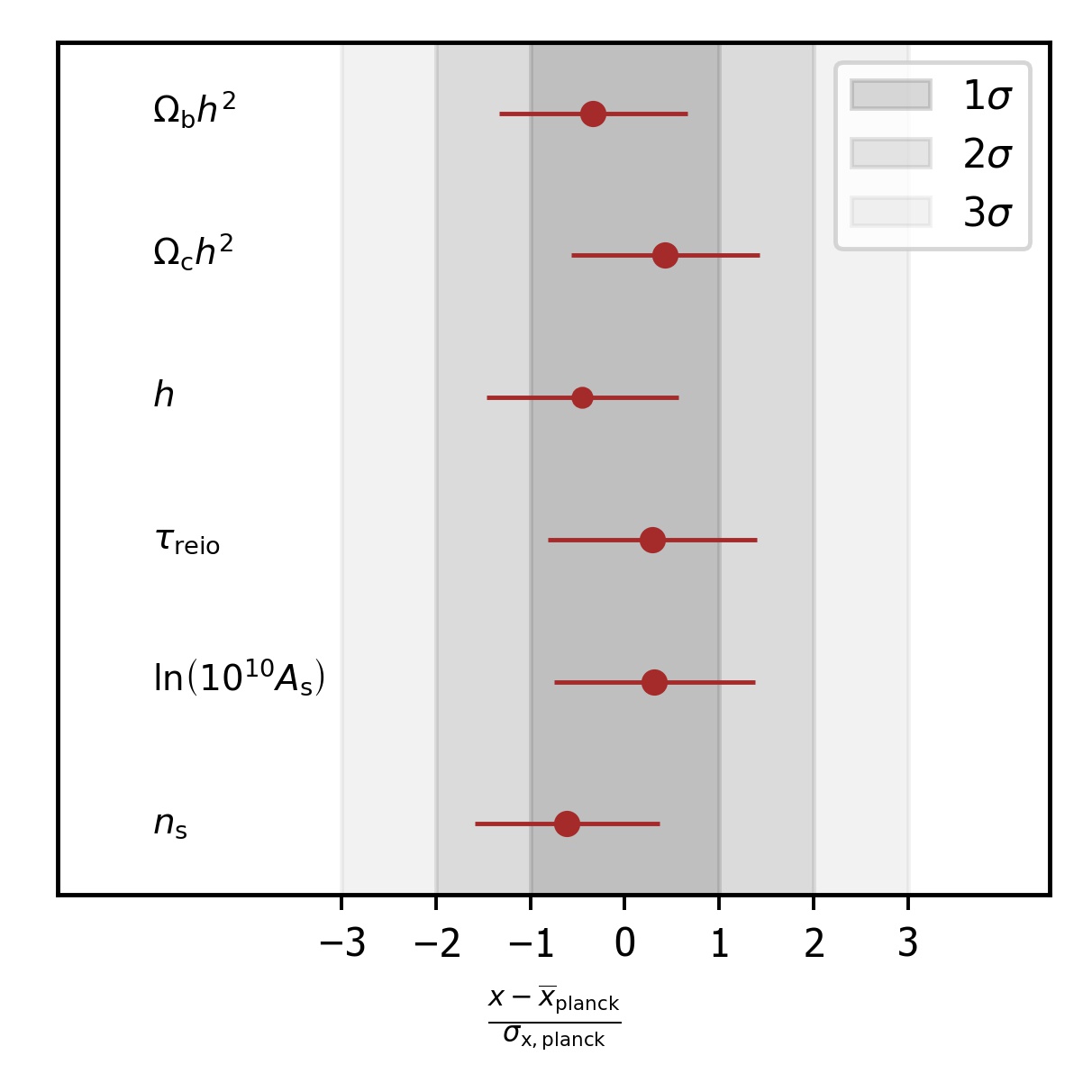}
    \caption{Values and $1\sigma$ confidence intervals of cosmological parameters resulting from our $\Lambda$CDM analysis related to the values obtained by \citet[P20b]{planck_2018} (TT,TE,EE+lowE). We also guide the eye by depicting the 1, 2, and $3\sigma$ intervals as grey bands.}
    \label{fig:lcdm_planck_benchmark}
\end{figure}

In Fig.~\ref{fig:lcdm_planck_benchmark} we show a comparison of our $\Lambda$CDM results with CMB \textit{Planck 2018} data and compare them to the result published in \planck~(see Tab.~2, setup TT,TE,EE+lowE.  {We also included lowT at the top of this setup}). We display all six inferred cosmological parameters centred on Planck values and normalized by Planck $1\sigma$ confidence intervals, thus displaying $\left(x - \overline{x}_{\rm planck}\right)/\sigma_{\rm x,planck}$ for a parameter $x$. Most of the parameters agree to $\sim 0.1\sigma$  {. The largest discrepancy is observed for $n_s$ at the level of $0.6\sigma$}.

\section{MCMC results}
\label{app:mcmc_details}

We present the detailed results of our MCMC analyses in Tab.~\ref{tab:lcdm_results} ($\Lambda$CDM) and \ref{tab:dddm_results} (DDM). In the top part of the tables, we show cosmological, baryonic, and DDM parameters directly sampled during the MCMC. The middle part displays the derived $\Omega_{\rm m}, \sigma_8$ and $S_8$ values, and the bottom part is dedicated to the details about the MCMC statistics (priors, likelihoods, and $\chi^2$~values). A long dash indicates that a specific parameter is not relevant for a specific setup, and \textit{unconst}  indicates unconstrained parameters.

\begin{table*}[!htb]
    \tiny
    \renewcommand{\arraystretch}{1.3}
    \centering
    \begin{tabular}{l c c c}
    \hline
    \hline
    \rule{0em}{-0.5em}\\
         & KiDS $\rm \Lambda CDM$ & Planck $\rm  \Lambda CDM$ & KiDS + Planck $\rm \Lambda CDM$\\
    Parameter     & 68\% limits  & 68\% limits & 68\% limits \\
    \hline
$\omega_{\rm c}$        &       $       0.146 (0.081)^{+0.034}_{-0.056} $       &       $       0.1208 (0.1200) \pm 0.0014     $       &       $       0.1182 ( 0.1188) \pm 0.0012     $        \\ 
$\omega_{\rm b}$        &       $        \rm unconst (0.02455)  $       &       $       0.02231 (0.02236) \pm 0.00015   $       &       $       0.02248 (0.02253) \pm 0.00013   $       \\ 
$\ln (10^{10} A_{\rm s})$       &       $        2.56 (3.50)^{+0.61}_{-0.80}    $       &       $       3.050 (3.110) \pm 0.017       $       &       $       3.039 (3.069)\pm 0.017  $       \\ 
$h_0$   &       $        \rm unconst (0.6041)   $       &       $       0.6700 (0.6734) \pm 0.0061     $       &       $       0.6815 (0.6799)^{+0.0050}_{-0.0057}     $       \\
$n_{\rm s}$             &       $        \rm unconst (0.9377)   $       &       $       0.9622 (0.9640) \pm 0.0043     $       &       $       0.9678 (0.9668)\pm 0.0041       $       \\ 
$\tau_{\rm reio}$       &       $       -       $       &       $       0.0566 (0.0841) \pm 0.0083     $       &       $       0.0540 (0.0677) \pm 0.0077      $       \\ 
$A_{\rm IA}$            &       $       0.75 (0.89)^{+0.33}_{-0.38}     $       &       $       -         $       &       $       0.63 (0.44)^{+0.25}_{-0.31}     $       \\ 
$A_{\rm planck}$        &       $       -       $       &       $       1.0005 (1.0031) \pm 0.0025     $       &       $       1.0003 (1.0005) \pm 0.0025      $       \\ 
$\log_{10} M_{\rm c}$           &       $       <13.1 (12.6)    $       &       $       -       $       &       $        >13.8 (15.0)    $       \\ 
$\theta_{\rm ej}$       &       $       <5.45 (2.23)    $       &       $       -       $       &       $        >5.88 (7.74)    $        \\ 
$\eta_\delta$           &       $        \rm unconst (0.21)     $       &       $       -       $       &       $        \rm unconst (0.13)      $       \\ 
 {$\log_{10} (\Gamma\times {\rm Gyr})$} &       $       -       $       &       $       -       $       &       $       -       $       \\ 
$f$     &       $       -       $       &       $       -       $       &       $       -       $       \\ 
\hline                                                                                                  
$\Omega_{\rm m}$        &       $       0.347^{0.066}_{-0.11}   $       &       $       0.3189 \pm 0.0087      $       &       $       0.3043 \pm 0.0070       $       \\ 
$\sigma_8$      &       $        0.70^{+0.11}_{-0.13}   $       &       $       0.8154 \pm 0.0080      $       &       $       0.8029\pm 0.0073        $       \\ 
$S_8$   &       $       0.735^{+0.031}_{-0.024} $       &       $       0.841 \pm 0.017       $       &       $       0.809 \pm 0.014 $       \\
\hline                                                                                                  
$\ln$(prior)            &       $       5.99    $       &       $       7.88 (0.76)^{+1.2}_{-0.41}   $       &       $       13.9 (12.9)^{+1.0}_{-0.31}      $       \\ 
ln(${\rm lik}_{\rm WL}$)        &       $       -81.3 ( {-79.3})^{+1.2}_{-0.54} $       &       $       -       $       &       $       -83.5 (-83.3)^{+1.7}_{-1.0}   $       \\ 
ln(${\rm lik}_{\rm CMB}$)       &       $       -       $       &       $       -294.4 (-290.1)^{+2.0}_{-1.1}  $       &       $       -295.7 ( {-291.9})^{+2.8}_{-1.6}        $       \\ 
$\chi^2_{\rm min}$      &       $        {158.7}        $       &       $       580.2   $       &       $        {750.5} $       \\ 
\hline                                                                                                  
    \end{tabular}
    \caption{The $\Lambda$CDM results of our MCMC analysis. We separately report individual results based on WL (\textit{KiDS-1000}) and CMB (\textit{Planck 2018}) data alone, as well as values inferred from the combined MCMC chain. We show the mean (best-fit) values of the sampled (top) and derived (middle) parameters and the obtained prior, likelihood, and $\chi^2$~values (bottom). }
    \label{tab:lcdm_results}
\end{table*}

\begin{table*}
    \tiny
    \renewcommand{\arraystretch}{1.3}
    \centering
    \begin{tabular}{l c c c}
    \hline
    \hline
    \rule{0em}{-0.5em}\\
         & KiDS $\rm 1bDDM$ & Planck $\rm  1bDDM$ & KiDS + Planck $\rm 1bDDM$\\
    Parameter     & 68\% limits  & 68\% limits & 68\% limits \\
    \hline
$\omega_{\rm c}$        &       $       0.137 (0.099)^{+0.033}_{-0.063} $       &       $       0.1209 (0.1205) \pm 0.0014     $       &       $       0.1184 (0.1176) \pm 0.0011      $        \\ 
$\omega_{\rm b}$        &       $        \rm unconst (0.02205)  $       &       $       0.02230(0.02236) \pm 0.00015     $       &       $       0.02246 (0.02258) \pm 0.00014   $       \\ 
$\ln (10^{10} A_{\rm s})$       &       $        2.74 (3.08)^{+0.70}_{-0.91}    $       &       $       3.052 (3.100) \pm 0.017       $       &       $       3.039 (3.089) \pm 0.017 $       \\ 
$h_0$   &       $        \rm unconst (0.6074)   $       &       $       0.6705 (0.6719) \pm 0.0062     $       &       $       0.6812 (0.6849) \pm 0.0052      $       \\
$n_{\rm s}$             &       $        \rm unconst (0.9480)   $       &       $       0.9618 (0.9628) \pm 0.0045     $       &       $       0.9671 (0.9685) \pm 0.0041      $       \\ 
$\tau_{\rm reio}$       &       $       -       $       &       $       0.0572 (0.0818) \pm 0.0084     $       &       $       0.0536 (0.0803) \pm 0.0082      $       \\ 
$A_{\rm IA}$            &       $       0.73 (0.82) \pm 0.34    $       &       $       -         $       &       $       0.64 (0.51) \pm 0.27    $       \\ 
$A_{\rm planck}$        &       $       -       $       &       $       1.0005 (1.0001) \pm 0.0025     $       &       $       1.0004 (0.9994) \pm 0.0025      $       \\ 
$\log_{10} M_{\rm c}$           &       $       <13.0 (12.8)    $       &       $       -       $       &       $       >13.9 (14.76) $       \\ 
$\theta_{\rm ej}$       &       $       <5.64 (2.60)    $       &       $       -       $       &       $       >5.85 (7.67)  $        \\ 
$\eta_\delta$           &       $        \rm unconst (0.28)     $       &       $       -       $       &       $        \rm unconst (0.24)      $       \\ 
 {$\log_{10} (\Gamma\times {\rm Gyr})$} &       $       <-2.24 (-2.67)  $       &       $       <-2.76 (-2.67) $       &       $       <-2.68 (-2.89)  $       \\ 
$f$     &       $        \rm unconst (0.841)    $       &       $       <0.603 (0.116) $       &       $       <0.602 (0.361)  $       \\ 
\hline                                                                                                  
$\Omega_{\rm m}$        &       $       0.323^{+0.073}_{-0.13}  $       &       $       0.3189 \pm 0.0087      $       &       $       0.3022^{0.0083}_{-0.0073}       $       \\ 
$\sigma_8$      &       $       0.73^{+0.12}_{-0.15}    $       &       $       0.8154 \pm 0.0080      $       &       $       0.8020 \pm 0.0072       $       \\ 
$S_8$   &       $       0.723^{+0.041}_{-0.027} $       &       $       0.841 \pm 0.017       $       &       $       0.805 \pm 0.015 $       \\
\hline                                                                                                  
ln(prior)       &       $       4.94    $       &       $       15.4 (10.1)^{+1.3}_{-0.42}      $       &       $       12.8 (7.8)^{+1.0}_{-0.40}    $       \\ 
ln(${\rm lik}_{\rm WL}$)        &       $       -81.3 (-79.4)^{+1.2}_{-0.59}    $       &       $       -       $       &       $       -83.6 (-82.91)^{+1.7}_{-0.95} $       \\ 
ln(${\rm lik}_{\rm CMB}$)       &       $       -       $       &       $       -294.5(-290.1)^{+2.1}_{-1.2}    $       &       $       -296.5 (-292.2)^{+2.6}_{-1.7}  $       \\ 
$\chi^2_{\rm min}$      &       $       158.7   $       &       $       580.1   $       &       $       750.2   $       \\ 
\hline                                                                                                  
    \end{tabular}
     \caption{The DDM results of our MCMC analysis. We separately report individual results based on WL (\textit{KiDS-1000}) and CMB (\textit{Planck 2018}) data alone, as well as values inferred from the combined MCMC chain. We show the mean (best fit) values of the sampled (top) and derived (middle) parameters and the obtained prior, likelihood, and $\chi^2$~values (bottom). }
    \label{tab:dddm_results}
\end{table*}

\end{document}